\documentstyle[12pt]{article}
\normalsize

\def\br{\begin{thebibliography}{abc}}
\def\er{\end{thebibliography}}
\def\rf{\bibitem}
\newcommand{\bc}{\begin{center}}
\newcommand{\ec}{\end{center}}
\begin{document}
\newcommand{\be}{\begin{equation}}
\newcommand{\ee}{\end{equation}}
\newcommand{\bea}{\begin{eqnarray}}
\newcommand{\eea}{\end{eqnarray}}


\baselineskip=24pt




\begin{center}\bf GROWTH OF METASTABLE HIGH-ORDER COMMENSURATE OVERLAYERS
OF Pb ON Cu(001) \end{center}

Wei Li and Gianfranco Vidali

Department of Physics, Syracuse University, Syracuse, NY 13244-1130 USA
\bigskip

\noindent ABSTRACT

\medskip\noindent
We studied the growth and ordering of a Pb layer deposited on Cu(001) at 150 K.
Contrary to the case of adsorption of Pb at room temperature, islands readily
form. These islands order in a high-order commensurate structure of symmetry
\mbox{($\sqrt{61}\times\sqrt{61}$)Rtan$^{-1}\left(\frac{5}{6}\right)$}. We
followed the growth of these
islands using atom beam scattering. From the measurement of the island
diffraction peak profiles we found a power growth law for the mean island- size
versus coverage.  Upon increasing the coverage another high-order commensurate
structure is formed, (5$\times$5)Rtan$^{-1}\left(\frac{3}{4}\right)$.
All the ordered phases obtained by adsorption at low temperature are
metastable.
\break
\newpage
\noindent
I. INTRODUCTION

The study of growth and ordering of overlayers deposited on cold single crystal
substrates poses special challenges. Recent experimental \cite{one,two} and
computational work \cite{3} on growth of metal overlayers on metal substrates
has
shown that our understanding of the interplay of kinetic, dynamic and energetic
processes during growth is inadequate. Of the many  metal-on metal systems
\cite{four},
particularly interesting is the study of lattice mismatched combinations.  In
these systems, ordered layers are the results of a competition of entropic
terms as well as of adsorbate-substrate and adsorbate-adsorbate interactions.
The absence of strongly directional bonds and of well tested interaction
potentials makes it difficult to carry out reliable calculations about
epitaxial processes.

Within this contest, we found that the study of the growth and ordered phases
of Pb on Cu(001) at low temperature can yield some valuable insights and a few
surprises as well. This is a lattice mismatched system (bulk lattice constant
of Cu=3.6 $\AA$, 4.5 $\AA$ for Pb) which has been extensively investigated in
the
past, but only for depositions on substrates at room temperature or higher (
410 K) \cite{five}.
A preliminary report \cite{two} of a study of adsorption of Pb at 150 K showed
the
appearance, near one layer coverage, of a high-order  commensurate structure
with 16 atoms per unit supercell (A unit supercell is the smallest cell
commensurate with the lattice). An intriguing observation about this phase
was that the proposed structure of the Pb layer had square symmetry
(The supercell has also square symmetry). This is surprising, since it is
usually believed that, for this type of metal-on-metal systems, high-order
commensurate or incommensurate phases prefer hexagonal or quasi-hexagonal
symmetries, since these have the highest coordination number \cite{six}.

Finally, we mention the work of Zuo and Wendelken \cite{seven}; investigating
the growth
of
ordered islands of Ag on Si(111) with High Resolution LEED (HRLEED), they found
that the size distribution of island is self-similar. From an analysis of peak
profiles, they found that the mean island size grows with increasing coverage
as $\Theta^{n}$. The value of the exponent n (0.20 to 0.35) was found dependent
of the
substrate temperature during deposition (350 to 450 $^{\circ}$C).

\bigskip\noindent
II. EXPERIMENTAL

The apparatus consists of a helium beam scattering line (18.4 meV incident
energy, velocity resolution $\delta$v/v $\sim$ 1$\%$ at 1200 PSI) coupled to a
UHV
chamber \cite{eight}. In the UHV chamber there are located: a liquid nitrogen
cooled
Knudsen evaporation source, a 4-grid LEED-Auger optics and a helium beam
detector, which is a differentially pumped quadrupole mass spectrometer with an
aperture of 0.5$^{\circ}$. This spectrometer can be rotated around the axis of
the sample manipulator and can be positioned to measure the reflected helium
beam during Pb deposition \cite{eight}. The deposition time for one monolayer
of Pb was
typically about 38 minutes. The coverage calibration was obtained from Auger
experiment as well as an analysis of the high temperature ordered phases.
Shorter deposition times did not affect the results.
Although the layers deposited at 150 K were found to be metastable (see below),
there was no indication from ABS, LEED or Auger of a rearrangement within the
layer during measurements.

The thoroughly desulfurized copper sample was cleaned prior to each run by Ar
ion sputtering and annealing at 580$^{\circ}$C.  Helium beam scattering was
used to assess the quality of the prepared surface as in previous
studies \cite{two,eight}.

\bigskip\noindent
III. RESULTS

In Fig.1  we present a sketch of the phase diagram. The phases below
0$^{\circ}$C are
metastable in the sense that they are obtained by depositing Pb at 150 K and
cannot be reached by cooling the overlayer deposited at room temperature. In
Fig.2 the LEED pattern and proposed structure of
($\sqrt{61}\times\sqrt{61}$)Rtan$^{-1}\left(\frac{5}{6}\right)$ are presented.
The ($\sqrt{61}\times\sqrt{61}$)Rtan$^{-1}\left(\frac{5}{6}\right)$
structure, with an ideal coverage of 0.49, was deduced from
He beam scattering data and from coverage considerations \cite{nine}.
The (5$\times$5)Rtan$^{-1}\left(\frac{3}{4}\right)$
structure has been shown elsewhere \cite{two,nine-2}.

At low coverage, Pb aggregates in islands with a
($\sqrt{61}\times\sqrt{61}$)Rtan$^{-1}\left(\frac{5}{6}\right)$
supercell; the evolution of the specular and diffraction peak profiles gives
information on how islands grow. Each diffraction peak was fitted with the
convolution of the instrument response function (well approximated by a
Gaussian) and a fitting function.
Following analyses of island growth studied by HRLEED, we tried Gaussian,
Lorentzian and power Lorentzian as fitting functions \cite{six,ten}.
In Fig.3, fits using a power Lorentzian:
\be
I(\Delta K_{\parallel}) \propto \frac{1}{(\xi^{2}+\Delta K_{\parallel}^2)^m}
\label{eq:lor}
\ee
\noindent
with m=5 are presented for three different coverages.
The same fitting function with m=5 was used for all coverages, from about 0.1ML
to 0.5ML. The
inverse width of the deconvolved peak (FWHM of Eq.(1)) is displayed as a
function of
coverage in Fig.4. The growth of the mean island size $\overline{R}$ versus
coverage can be
well described by the equation:
\be
\overline{R} \sim \frac{1}{FWHM} \propto \Theta ^n \label{eq:cov}
\ee

\smallskip\noindent
The average size of ordered islands goes from about 30$\AA$ to 80$\AA$.
We also performed an independent analysis of the diffraction beam shapes. Each
data
set was deconvolved of the instrument response function  using Fourier
analysis. Notice that these deconvolved data have been obtained without
assuming any functional form for the "true" signal. In Fig. 5 we superimposed
deconvolved data from three different coverages after scaling the horizontal
axis. The fact that their shape is the same indicates that growth is
self-similar in the range of coverage investigated. The oscillations in the
wings of the peaks are due to
the  deconvolution procedures; in fact, they are not present in the original
data, see Fig. 3.

\bigskip\noindent
IV. DISCUSSION

The two ordered phases obtained after deposition at 150K are metastable; upon
heating them above about 0$^{\circ}$C they convert to the phases obtained by
depositing
Pb at 400 K. These latter ones are equilibrium phases, since they can be melted
and recrystallized \cite{two,nine}. The metastable phases have quite a
different structure
 than
the equilibrium ones, since they are in registry with the substrate via very
large supercells. Only one atom in 16 for the
(5$\times$5)Rtan$^{-1}\left(\frac{3}{4}\right)$
or 30 for the ($\sqrt{61}\times\sqrt{61}$)Rtan$^{-1}\left(\frac{5}{6}\right)$
is in registry of the substrate,
indicating that they are the results of limited diffusion and of a delicate
competition between Pb-Pb and Pb-Cu interactions. It is unusual to find
structures with so few atoms in registry with the substrate; in fact, in most
cases the overlayer structure becomes incommensurate with a hexagonal unit
cell. A calculation of the energetics of these phases is in progress and will
be reported elsewhere.

We found that the distribution of island size is self-similar, see Fig.5, that
is the peak profiles can be superimposed after a rescaling of the horizontal
axis.
For the case of growth of two-dimensional islands, we are not aware of any
theory that makes a prediction about
the exponent in Eq.(2) \cite{eleven}. It is not known the shape of the
universal
function describing the distribution of island sizes; Eq.(1) should be
considered at this stage as a convenient fitting function. The value of
the exponent  we obtained is different from the ones of Ref. 7. In that case,
the highest exponent, 0.35, has been obtained for the highest substrate
temperature, 450 $^{\circ}$C. At this stage it is too premature to speculate
whether the
values of n can be grouped in universality classes. The temperature dependence
of n might indicate that different processes are activated at different
temperatures. Because of technical limitations we are not able at the present
time to obtain information on the temperature dependence of n for our system.

\noindent
ACKNOWLEDGMENTS

This work was supported by NSF grant DMR 8802512.

\newpage

\noindent
FIGURE CAPTIONS

\noindent
Figure 1.  Sketch of phase diagram of Pb on Cu(001) obtained by ABS and LEED
data. Crossed region: disordered; cross-hatched region: phases coexist.

\noindent
Figure 2.  Reciprocal space (left) and proposed real space structure (right) of
($\sqrt{61}\times\sqrt{61}$)Rtan$^{-1}\left(\frac{5}{6}\right)$.

\noindent
Figure 3.  (0,-1) ABS diffraction peak from
($\sqrt{61}\times\sqrt{61}$)Rtan$^{-1}\left(\frac{5}{6}\right)$ islands at
representative coverages; Solid lines are the fitting results of the
convolution of the instrument response function
with the function in Eq.(1). $\theta_i$=60 $^{\circ}$.

\noindent
Figure 4.  Inverse FWHM vs. coverage obtained from an analysis of diffraction
peaks as in Fig.3. The line is the best-fit through the data, see Eq.(2).

\noindent
Figure 5.  ABS diffraction data from islands after deconvolution of the
instrument response function. Data from
three coverages have been rescaled and plot in the same  graph.
$\theta_i$=60$^{\circ}$.  w is the peak full width at half maximum (FWHM).

\end{document}